\documentclass[prb,aps,twocolumn,superscriptaddress,showpacs]{revtex4-1}
\usepackage{graphicx,color}
\usepackage{amsthm}
\usepackage{amsfonts}
\usepackage{algorithmic}
\usepackage{enumerate}
\usepackage{latexsym}
\usepackage{amsmath}
\usepackage{amssymb}
\usepackage{bm}
\usepackage[pdftex,plainpages=false,colorlinks=true,linkcolor=blue, citecolor=blue, urlcolor=blue]{hyperref}

\emergencystretch=\maxdimen
\begin{document}
\title{Phase diagram of the Hubbard model on a honeycomb lattice: \\A cluster slave-spin study}

\author{Ming-Huan Zeng}
%\email{minghuanz@mail.bnu.edu.cn}
\affiliation{Department of Physics, Beijing Normal University, Beijing 100875, China}
\author{Y.-J. Wang}
\email{yjwang@bnu.edu.cn}
\affiliation{Department of Physics, Beijing Normal University, Beijing 100875, China}
\author{Tianxing Ma}
\email{txma@bnu.edu.cn}
\affiliation{Department of Physics, Beijing Normal University, Beijing 100875, China}

\date{\today}

\begin{abstract}
The cluster slave-spin method is implemented to research the ground state properties of the honeycomb lattice Hubbard model
with doping $\delta$ and coupling $U$ being its parameters. At half-filling, a single direct and continuous phase
transition between the semi-metal and antiferromagnetic (AFM) insulator is found at $U_{\text{AFM}}=2.43t$ that is in the
Gross-Neveu-Yukawa universality class, where a relation between the staggered magnetization $M$ and the AFM energy gap
$\Delta_{\text{AFM}}$ is established as $M \propto \Delta_{\text{AFM}}$, compared to $M \propto \Delta_{\text{AFM}}
( \ln{\Delta_{\text{AFM}}})^2$ in the square lattice case.  A first-order semi-metal to the underlying paramagnetic (PM)
insulator Mott transition is corroborated at $U_{\text{Mott}}=8.36t$, which is responsible for a broad crossover around
$U_{c} = 5.4t$ between the weak- and strong-coupling regimes in the AFM state that increases with $\delta$,
in contrast to the square lattice case. In the doped system, the compressibility $\kappa$ near the van Hove
singularity at $\delta=1/4$ is suppressed substantially by the interaction before the semi-metal to AFM transition occurs,
whereas $\kappa$ near the Dirac points is very close to the noninteracting one, indicating that the Dirac cone structure
of the energy dispersion is rather robust. An overall phase diagram in the $U$-$\delta$ plane is presented, consisting of
four regimes: the AFM insulator at $\delta=0$ for $U> U_{\text{AFM}}$, the AFM metal with compressibility $\kappa>0$
or $\kappa<0$, and the PM semi-metal, and the AFM metal with $\kappa<0$ only exists in an extremely small area
near the phase boundary between the AFM and PM state.
\end{abstract}


\maketitle

%%%%%%%%%%%%%%%%%%%%%%%%%%
%%%%%%%%%%%%%%%%%%%%%%%%%%
\section{INTRODUCTION}
%%%%%%%%%%%%%%%%%%%%%%%%%%
%%%%%%%%%%%%%%%%%%%%%%%%%%

The Hubbard Hamiltonian~\cite{Hubbard1963,*Hubbard1964} has been acting over decades as a prototypical model for
the description of interacting electrons. In spite of its seeming simplicity, this model captures a rich phenomenology
of strongly-correlated electrons such as metallic-insulating, nonmagnetic-antiferromagnetic and normal-superconducting
phase transitions and can not be solved exactly in more than one dimensions, which necessitates some nonperturbative
approaches to deal with the strong-coupling aspect of the model~\cite{PhysRevB.40.506,Mingpuqin2021}. In this paper,
we will focus on the one-band Hubbard model defined on a honeycomb (hexagon) network (Bravais lattice for
graphene), which is bipartite and admits the antiferromagnetism (AFM) in the strong coupling limit. This model has
a linear free electron energy dispersion with nodal gapless points at the corners of the Brillouin zone, leading to
the so-called Dirac semi-metal. Due to the gapless Dirac points, there is a nontrivial semi-metal to antiferromagnetic
insulator (AFMI) transition at a finite coupling strength at half-filling, which makes this model an ideal playground
to research the interaction-driven semi-metal to AFMI transition. Up till now, many numerical and analytical methods
have been applied to the half-filled system to study this transition and its critical behaviors. Large scale quantum
Monte Carlo (QMC) simulations of $648$ sites predict a spin liquid state in a range of interaction $3.5t<U<4.3t$,
beyond which the AFM sets in~\cite{Meng_2010}, and this argument was supported by some numerical
works~\cite{PhysRevLett.106.100403, PhysRevLett.109.229902,PhysRevB.85.115132,PhysRevB.84.205121}. Nevertheless, this
picture was disputed by many other numerical studies~\cite{Otsuka_2016, Raczkowski_2020, Assaad_2013, Paiva_2005,
Yamada_2016, Sorella_1992, Sorella_2012, Ostmeyer_2020,ostmeyer_2021}, especially those using the same method
containing up to $2592$ sites~\cite{Sorella_2012, Otsuka_2016} and $20808$ sites~\cite{Ostmeyer_2020, ostmeyer_2021}.
By means of cluster dynamical mean-field theory, variational cluster approximation, and cluster dynamical impurity
approximation, Hassan \emph{et al.}~\cite{Hassan_2013} showed that the results are dependent on the shape and size
of the clusters, and they claimed that only the system with two bath orbits per cluster boundary site is able to
describe the correct behavior and found that the Mott transition for the spin liquid state is actually preempted by
the AFM long-range order. Though the early variational cluster calculations~\cite{seki_2012} argued that the
single-particle gap opens at an infinitesimal value of $U$, recent dynamical cluster approximation study found
that this spurious excitation gap is due to the violation of the translation symmetry of the system and the cluster
with one bath orbital per cluster site is sufficient for the description of the short-range correlations within the
honeycomb unit cell~\cite{Liebsch_2013}. A recent density matrix embedding theory study revealed a paramagnetic
insulating state with possible hexagonal cluster state at intermediate coupling strength whose stability is highly
cluster and lattice size dependent, and this state is nonexistent in the thermodynamic limit, signaling no intermediate
state in the half-filled Hubbard model on a honeycomb lattice~\cite{Chen_2014}. In addition, a two-particle
self-consistent study presented a semi-metal to AFMI transition and proved that the transition from a semi-metal to
spin liquid phase is forestalled by this transition~\cite{Arya_2015}. The functional renormalization group theory
predicts a critical interaction strength $U=3.8t$ that is consistent with the results from the methods mentioned
above, supporting that there is no spin liquid state at intermediate coupling
strengths~\cite{Honerkamp_2008, Raghu_2008}.

Based on the charge-spin separation theory~\cite{Y.R.Wang_1993, Feng_1993, *Feng_1994, *Feng_2003, *Feng_2004,
*Feng_2015, PhysRevB.81.035106}, the $U(1)$ slave-spin method has been proposed to cope with the Mott transition
in multi-orbital systems~\cite{Yu.Rong_2012}, which is very economical computationally because only
$2M$ slave spins need to be introduced per site with $M$ being the number of orbits. This method
can not only reproduce the Gutzwiller factor $g_t = \frac{1 - x - 2d}{1 - x - (1 - x)^{2}/2}
(\sqrt{x + d} + \sqrt{d})^{2}$, but also capture the right noninteracting behaviors at $U = 0$
because of an extra orbital-dependent chemical potential in the spinon Hamiltonian,
which makes it a powerful method to deal with the strong-coupling systems~\cite{Lee.Ting-Kuo_2017}.
Then, a cluster slave-spin approach was developed to address strongly correlated systems to take the short-range
charge fluctuations into account~\cite{Lee.Ting-Kuo_2017}
and has been employed to solve the square lattice Hubbard model to obtain
an overall ground state phase diagram in the parameter space of doping $\delta$ and interaction $U$~\cite{zeng2021}.
In the present work, we apply the same method with the Lanczos exact-diagonalization as the slave-spin cluster solver
to the honeycomb lattice Hubbard model to study its ground state properties, including the quantum critical behavior
in the vicinity of the interaction-driven semi-metal to AFMI transition at half-filling  and an overall phase diagram
in the whole $U$-$\delta$ plane.
Our motivation is two-fold:
(i) Because the results of this model is shown to be highly dependent on the size of the lattice adopted for QMC
simulations, as well as the size and shape of the clusters used within various cluster approximations, more results
from different approaches ought to be included and compared with each other.
(ii) Away from half-filling, much attention was paid to the 1/4-doping, where the free density of states shows a van
Hove singularity of logarithmic type, favoring an instability towards superconductivity in the weak interaction
regime~\cite{PhysRevB.88.155112, PhysRevB.85.035414, Nandkishore2012}, whereas an overall $U$-$\delta$ phase diagram
pertaining to the magnetism is still absent.

In the honeycomb lattice Hubbard model, we find that the first-order Mott transition occurs at $U_{\text{Mott}}=8.36t$
in the half-filled PM state, characterized by discontinuities and hystereses in all quantities, and transforms into
a broad crossover in the AFM state because of long range AFM correlations. Besides, the phase separation, manifested
by a negative compressibility, has been observed in a region near the phase boundary $\delta_{M}(U)$ between the AFM
and PM state and at intermediate couplings, whose area is much smaller compared to the square lattice Hubbard
model~\cite{zeng2021}. Finally, a phase diagram in the $U$-$\delta$ plane is presented, consisting of four regimes:
AFMI, AFM metal with positive and negative compressibility, and the PM semi-metal.

The rest of this paper is organized as follows.
In Sec.~\ref{S-SPINFORMULA}, we reintroduce the cluster slave-spin mean-field theory~\cite{Lee.Ting-Kuo_2017,zeng2021}
and implement it in  the honeycomb lattice Hubbard model by making use of two- and six-site cluster approximations.
In Sec.~\ref{HALF-FILL}, for the half-filled system, an analytical relation between the staggered magnetization $M$
and the AFM energy gap $\Delta_{\text{AFM}}$ in the vicinity of the semi-metal to AFMI transition is established,
and the first-order Mott transition at $U_{\text{Mott}}=8.36t$ is observed in the PM state.
In Sec.~\ref{FINITEDOPE},  the results of finite doping cases obtained by two- and six-site clusters are discussed
thoroughly, and we find that the two-site cluster is inadequate to capture the AFM transition appropriately because
it violates the symmetry of the honeycomb lattice. In Sec.~\ref{PHASES}, the properties of $M$,
$\Delta_{\text{AFM}}$, and the compressibility $\kappa$ are combined to show a phase diagram
of the model in the $U$-$\delta$ plane.

%%%%%%%%%%%%%%%%%%%%%%%%%%
%%%%%%%%%%%%%%%%%%%%%%%%%%
\section{Formalism}\label{S-SPINFORMULA}
%%%%%%%%%%%%%%%%%%%%%%%%%%
%%%%%%%%%%%%%%%%%%%%%%%%%%

The standard one-band fermionic Hubbard model~\cite{Hubbard1963,Hubbard1964} reads
\begin{equation}\label{Hubbard model}
H = -t \sum_{\langle i,j \rangle \sigma} (c_{i\sigma}^\dagger c_{j\sigma} + \mbox{\sc h.c.}) + U \sum_{i}
n_{i\uparrow} n_{i\downarrow}
- \mu \sum_{i\sigma} n_{i\sigma} \;,
\end{equation}
where $t$, $U$, $\mu$ are the nearest hopping constant, the on-site Coulomb repulsion energy and the chemical
potential, respectively. The sum $\langle i,j \rangle $  runs over all pairs of nearest-neighbor sites on a honeycomb
lattice, and $c_{i\sigma}^\dagger$ is the creation operator of the electron at site $i$ with spin $\sigma= \uparrow,\,
\downarrow$, and the number operator $n_{i\sigma} = c_{i\sigma}^\dagger c_{i\sigma}$.  Hereafter, we use $t=1$ as the
unit of energy.

In the U(1)  slave-spin method~\cite{Yu.Rong_2012}, an electron  operator is factorized into a slave-spin operator
($S = \tfrac{1}{2}$) and a fermionic spinon operater, describing the charge and spin degrees of freedoms of an
electron, respectively:
\begin{equation}\label{s-spin-representation}
c^{\dagger}_{\alpha}\equiv S^{\dagger}_{\alpha}f^{\dagger}_{\alpha}\;,
\end{equation}
on account of which the original Hillbert space with basis
$\{|0\rangle, |1\rangle\}$ is enlarged to $\{|n^{f}_{\alpha},S^{z}_{\alpha}\rangle\} =
\{|0,-\tfrac{1}{2}\rangle,|1,\tfrac{1}{2}\rangle, |0,\tfrac{1}{2}\rangle,|1,-\tfrac{1}{2}\rangle\}$.
Thus, an extra constraint needs to be imposed to restrict the Hillbert space to the physical one:
$\{|n^{f}_{\alpha},S^{z}_{\alpha}\rangle\}=
\{|0,-\tfrac{1}{2}\rangle,|1,\tfrac{1}{2}\rangle\}$,
\begin{equation}\label{constraint}
S^{z}_{\alpha}=f^{\dagger}_{\alpha}f_{\alpha}-\tfrac{1}{2}\;.
\end{equation}
A gauge degree of freedom must be introduced to incorporate the constraint, signifying that the slave-spin
representation is invariant under a local gauge transformation $f^{\dagger}_{\alpha} \to f^{\dagger}_{\alpha}
e^{-i \phi_{\alpha}}$ and $S^{\dagger}_{\alpha} \to S^{\dagger}_{\alpha} e^{i \phi_{\alpha}}$, and all
physical quantities should be invariant under this U(1) gauge transformation~\cite{Coleman_1987,PatrickA_1992,Feng_1993,Serge_2004,Senthil_2008}.

With the constraint $a^{\dagger}_{\alpha}a_{\alpha}+b^{\dagger}_{\alpha}b_{\alpha}=1$, the slave-spin operator
is rewritten in the  Schwinger boson representation
\begin{equation}
S^{\dagger}_{\alpha} = a^{\dagger}_{\alpha}b_{\alpha} \;, \;\;\;\;
S^{z}_{\alpha} = \frac{1}{2}\left(a^{\dagger}_{\alpha}a_{\alpha}-b^{\dagger}_{\alpha}b_{\alpha}\right)\;.
\end{equation}

To ensure the correct non-interacting behaviors, the slave-boson operators need to be dressed as
follows~\cite{Kotliar_1986}
\begin{subequations}
\begin{eqnarray}
\tilde{S}^{\dagger}_{\alpha} &=& P^{+}_{\alpha}a^{\dagger}_{\alpha}b_{\alpha}P^{-}_{\alpha}\;, \\
P^{\pm}_{\alpha} &=& \frac{1}{\sqrt{1/2 \pm S^{z}_{\alpha}}}\;,
\end{eqnarray}
\end{subequations}
which can be linearized as follows:
\begin{equation}
\tilde{S}^{\dagger}_{\alpha} \approx \tilde{z}^\dagger_{\alpha}
+ \frac{\langle\tilde{z}^\dagger_{\alpha}\rangle \langle S^z_{\alpha} \rangle \Delta S^z_{\alpha}}
{( \tfrac{1}{2} )^2 - \langle S^z_{\alpha} \rangle^2 }\;,
\end{equation}
with $\Delta S^z_{\alpha} = S^z_{\alpha} - \langle S^z_{\alpha} \rangle$
and $\tilde{z}^\dagger_{\alpha} =  a^\dagger_{\alpha} b_{\alpha} /
[ (\tfrac{1}{2})^2 - \langle S^z_{\alpha} \rangle^2 ]^{1/2}$.

Following the recipe of Lee and Lee~\cite{Lee.Ting-Kuo_2017}, with the local constraints~\eqref{constraint} being
ensured roughly by two global
Lagrange multipliers $\lambda_{I\sigma}$ on sublattices $I=A$ and $B$, Hamiltonian \eqref{Hubbard model} can be
cast into the form
\begin{subequations}\label{HAMILTONIAN}
\begin{eqnarray}
H^f_{\text{MF}}& =&  -tZ\sum_{\langle i,j\rangle \sigma} (a_{i\sigma}^\dagger b_{j\sigma}
+ \textsc{h.c.}) \nonumber\\
&&-\sum_{i\sigma}\big[(\mu+\lambda_{A\sigma} -\tilde{\mu}_{A\sigma})a_{i\sigma}^\dagger a_{i\sigma} \nonumber\\
&&\hspace{2.3em} +(\mu+\lambda_{B\sigma} -\tilde{\mu}_{B\sigma})b_{i\sigma}^\dagger b_{i\sigma}
\big] \;, \label{FERMI-HAML}\\
H_{n_c\text{-site}}^{S} &=& H^\lambda_{n_c\text{-site}}+H^U_{n_c\text{-site}}+H^K_{n_c\text{-site}}
\;,\label{SSPIN-HAML-nc-SITE}
\end{eqnarray}
\end{subequations}
where
\begin{subequations}\label{SSPIN-H}
\begin{eqnarray}
H^\lambda_{n_c\text{-site}} &=& \sum_{i_c=1,\sigma}^{n_c}\lambda_{I\sigma}S_{i_c\sigma}^{z}\;,\\
H^U_{n_c\text{-site}} &=& \sum_{i_c=1}^{n_c}U(S_{i_c\sigma}^{z}+\tfrac{1}{2})(S_{i_c\bar{\sigma}}^{z}
+\tfrac{1}{2})\;,\\
H^K_{\text{2-site}} &=& \sum_{\sigma} \Big\{ \epsilon_{\sigma}^{\delta_1} ( \tilde{z}_{A\sigma}^\dagger
\tilde{z}_{B\sigma} +\tilde{z}_{B\sigma}^\dagger \tilde{z}_{A\sigma} )
+ (\epsilon_{\sigma}^{\delta_2} +  \epsilon_{\sigma}^{\delta_3} ) \nonumber\\
&& \hspace{2em} \times \big[ \tilde{z}_{A\sigma}^\dagger \langle \tilde{z}_{B\sigma} \rangle
+ \tilde{z}_{B\sigma}^\dagger \langle \tilde{z}_{A\sigma} \rangle + \textsc{h.c.} \big] \Big\} \;,\\
H^K_{\text{6-site}} &=& \sum_{\sigma}\Big\{ \epsilon_\sigma^{\delta_1}( \tilde{z}_{1\sigma}^\dagger
\tilde{z}_{2\sigma}+\tilde{z}_{4\sigma}^\dagger \tilde{z}_{5\sigma})\nonumber\\
&& +\epsilon_\sigma^{\delta_2}( \tilde{z}_{1\sigma}^\dagger \tilde{z}_{6\sigma}+\tilde{z}_{3\sigma}^\dagger
\tilde{z}_{4\sigma})
+\epsilon_\sigma^{\delta_3}( \tilde{z}_{2\sigma}^\dagger \tilde{z}_{3\sigma}+\tilde{z}_{5\sigma}^\dagger
\tilde{z}_{6\sigma})\nonumber\\
&&+\epsilon_{\sigma}^{\delta_1}(\tilde{z}_{3\sigma}^\dagger \langle \tilde{z}_{6\sigma}
\rangle+\langle\tilde{z}_{3\sigma}^\dagger\rangle \tilde{z}_{6\sigma}) \nonumber\\
&& +\epsilon_{\sigma}^{\delta_2}(\tilde{z}_{2\sigma}^\dagger \langle \tilde{z}_{5\sigma}
\rangle+\langle\tilde{z}_{2\sigma}^\dagger\rangle \tilde{z}_{5\sigma})\nonumber\\
&& +\epsilon_{\sigma}^{\delta_3}(\tilde{z}_{1\sigma}^\dagger \langle \tilde{z}_{4\sigma}
\rangle+\langle\tilde{z}_{1\sigma}^\dagger\rangle \tilde{z}_{4\sigma})+
\textsc{h.c.}\Big\} \;. \label{H_K}
\end{eqnarray}
\end{subequations}
The mean-field Hamiltonian $H_{n_c\text{-site}}^{S}$ is a Bose-Hubbard model for two species of bosons, and
actually a model of interacting XY spins in a magnetic field~\cite{Yu.Rong_2012}.
Senthil has systematically investigated the gauge field fluctuations' effects on charge
and spin degrees of freedom of the one-band Hubbard model in the slave-rotor
representation~\cite{Serge_2004}, which is very similar to the slave-spin method adopted in this paper. He found
that~\cite{Senthil_2008} the dynamical exponent $z = 1$ at the mean-field critical fixed point renders the Landau
damping term of the gauge bosons, $|\omega|/q$, scales as a Higgs mass term. Hence, for the rotors, the gauge bosons
are gapped and harmless, indicating that the universality class of the rotor quantum critical point remains unaltered
from 3D XY model~\cite{Witczak-Krempa_2013}. In this paper, the gauge fluctuations will not be considered further on the same ground.

\begin{figure}[htb!]
\centering{
\includegraphics[
width=0.46\textwidth%,scale=0.56
]{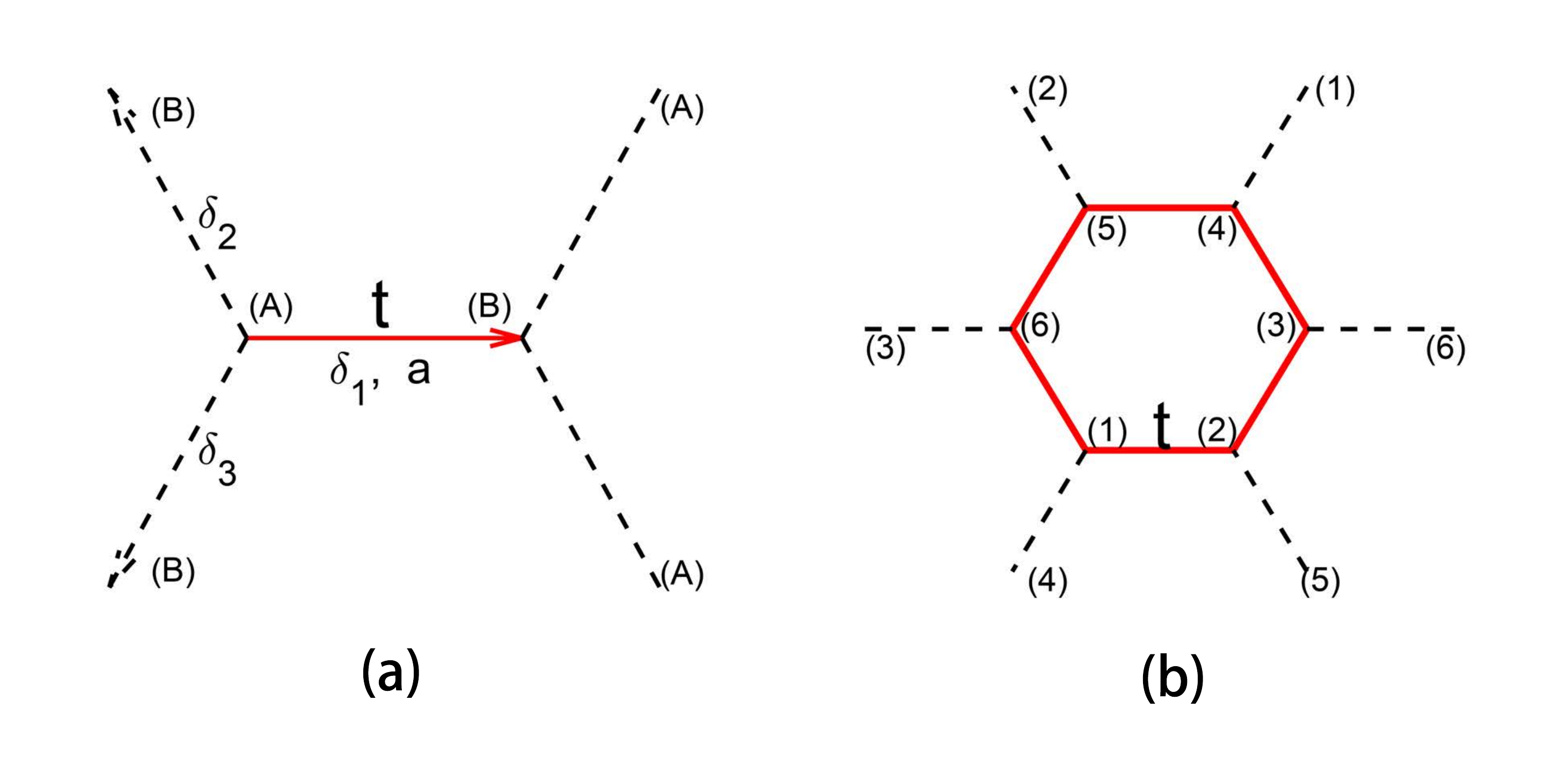}
}
\caption{Schematic illustration of the (a) two- and (b) six-site cluster configurations. The lattice constant
$a$ is set to unity, and the positions of three nearest neighbors of site A are $\delta_1=(1,0)$,
$\delta_2=(-1/2,\sqrt{3}/2)$, $\delta_3= -(1/2,\sqrt{3}/2)$.}
\label{twosixcluster}
\end{figure}

The cluster slave-spin Hamiltonian \eqref{SSPIN-HAML-nc-SITE} with $n_c =$ 2, 6,  marked by
the red color geometry in Fig.~\ref{twosixcluster}, will be solved by using the Lanczos exact-diagonalization
method. The parameters $Z$,
$\tilde{\mu}_{I\sigma}$, and $\epsilon_{\sigma}^{\delta}$ in Eqs.~\eqref{HAMILTONIAN}
and \eqref{SSPIN-H} are calculated as follows:
\begin{eqnarray}
Z&=&\langle\tilde{z}_{A\sigma}^\dagger\rangle\langle\tilde{z}_{B\sigma}\rangle,\;\hspace{2em}
\epsilon_{\sigma}^{\delta_{1/2/3}} =
-t \langle a_{i\sigma}^\dagger b_{i+\hat{\delta}_{1/2/3}\sigma} \rangle,\nonumber\\
&& \tilde{\mu}_{I\sigma} = \frac{2 Z \langle S_{I\sigma}^z \rangle (\epsilon_\sigma^{\delta_1}
+ \epsilon_\sigma^{\delta_2} +
\epsilon_\sigma^{\delta_3}) }{( \tfrac{1}{2} )^2 - \langle S_{I\sigma}^z \rangle^2 }\;.
\end{eqnarray}

Moreover, the fermionic spinon Hamiltonian can be Fourier transformed into momentum space:
\begin{eqnarray}
H^f_{\text{MF}}
&=& \sum_{\bm{k},\sigma} \big( \varepsilon_{A\sigma} a_{\bm{k}\sigma}^\dagger a_{\bm{k}\sigma}
+ \varepsilon_{B\sigma} b_{\bm{k}\sigma}^\dagger b_{\bm{k}\sigma} \nonumber \\
&& \hspace{2em} + \Gamma_{\bm{k}} a_{\bm{k}\sigma}^\dagger b_{\bm{k}\sigma}
+ \Gamma_{\bm{k}}^* b_{\bm{k}\sigma}^\dagger a_{\bm{k}\sigma} \big) \label{FERMI-HAML-K}
\end{eqnarray}
with
\begin{eqnarray}
&& \varepsilon_{A/B\sigma} = \tilde{\mu}_{A/B\sigma} -\mu -\lambda_{A/B\sigma} \;, \nonumber \\
&& \Gamma_{\bm{k}} = -tZ \gamma_{\bm{k}}\;, \;\;\;\; \gamma_{\bm{k}} = \sum_{\bm{\delta}} e^{i\bm{k}
\cdot \bm{\delta}} \;.
\end{eqnarray}
Diagnalization of the spinon Hamiltonian~\eqref{FERMI-HAML-K} gives rise to the eigenenergies as
\begin{subequations}
\begin{eqnarray}
E_{\bm{k}}^{\pm} &=& \pm \sqrt{|\Gamma_{\bm{k}}|^2 +\Delta_{\sigma}^2} -\mu_{\text{eff}}\;, \\
\mu_{\text{eff}} &=& \mu -\frac{1}{2}(\tilde{\mu}_{A\sigma} - \lambda_{A\sigma} + \tilde{\mu}_{B\sigma}
- \lambda_{B\sigma})\;, \\
\Delta_\sigma &=&\frac{1}{2}(\tilde{\mu}_{A\sigma} - \lambda_{A\sigma} - \tilde{\mu}_{B\sigma}
+ \lambda_{B\sigma})\;.
\end{eqnarray}
\end{subequations}
Here, AFM energy gap $\Delta_{\text{AFM}}=|\Delta_\sigma|$ is identical in form to that in the square lattice
case\cite{Lee.Ting-Kuo_2017,zeng2021}.

In most occasions, it proves effective to adopt the density of states (DOS) of the non-interacting electrons
to calculate the physical quantities in the thermodynamic limit.
On the honeycomb lattice, it is defined as
\begin{eqnarray}
D(\gamma) &=& \frac{1}{{\cal N}_\text{triangle}} \sum_{\bm{k}} \delta(\gamma - |\gamma_{\bm{k}}|) \nonumber \\
&=& \Big\{ \begin{array}{ll} N(\gamma) \;, &\;\;\; (0\le \gamma <1) \\
\tilde{N}(\gamma) \;, &\;\;\; (1< \gamma \le 3) \end{array} \label{density}
\end{eqnarray}
 where ${\cal N}_\text{triangle}$ is the site number of the underlying triangular lattice,
which is half of that of the honeycomb lattice, and
\begin{equation}
\Bigg\{ \begin{array}{l}
N(\gamma) = \frac{4}{\pi^2} \frac{\gamma}{\sqrt{(3- \gamma)(1 + \gamma)^3}}
\,K\Big(\sqrt{\frac{16\gamma} {(3- \gamma)(1 + \gamma)^3}} \,\Big) \;, \\
\tilde{N}(\gamma) = \frac{1}{\pi^2} \sqrt{\gamma}
\,K\Big(\sqrt{\frac{(3- \gamma)(1 + \gamma)^3} {16\gamma}} \,\Big) \;, \end{array}
\end{equation}
with $K(x)$ being the first kind complete elliptical integral.
In comparison to the self-dual situation on a square lattice, we now have a duality transformation
\begin{equation}
\tilde{\gamma} = \frac{3-\gamma}{1+\gamma}
\end{equation}
to connect these two parts, under which
\begin{subequations}
\begin{eqnarray}
\frac{(3- \tilde{\gamma})(1 + \tilde{\gamma})^3} {16\tilde{\gamma}}
&=& \frac {16\gamma}{(3- \gamma)(1 + \gamma)^3}\; ,\\
\frac{(3- \tilde{\gamma})(1 + \tilde{\gamma})} {4\tilde{\gamma}}
&=& \frac {4\gamma}{(3- \gamma)(1 + \gamma)}\;,
\end{eqnarray}
\end{subequations}
and
\begin{eqnarray}
\tilde{N}(\gamma) &=& \frac{(3-\tilde{\gamma})(1+\tilde{\gamma})}{4\tilde{\gamma}} N(\tilde{\gamma}) \nonumber \\
&=& \frac{4\gamma}{(3-\gamma)(1+\gamma)} N\Big(\frac{3-\gamma}{1+\gamma} \Big) \;.
\end{eqnarray}

Then, the self-consistent quantities $\epsilon^{\delta}_{\sigma}= \epsilon$ and $n_{(A/B)\sigma} \equiv \langle
a^\dagger_{i\sigma}a_{i\sigma}\rangle
 / \langle b^\dagger_{i\sigma}b_{i\sigma}\rangle$ can be calculated through
\begin{eqnarray}
&& \!\!\!\!\!\!\!\! \epsilon = \int_{0}^{3} \!\!d\gamma\, D(\gamma)\frac{(tZ\gamma)^2}{6Z \sqrt{(tZ\gamma)^2
+\Delta^2}} \sum_{s=\pm} s\theta[-E^s(\gamma)] \;, \label{EPS-F} \\
&& \!\!\!\!\!\!\!\! n_{(A/B)\sigma} = \int_{0}^{3} \!\!d\gamma\, D(\gamma) \sum_{s=\pm} \theta[-E^s(\gamma)] \nonumber \\
&& \hspace{4em} \times \frac{1}{2} \Big(1 \pm s \frac{\Delta_\sigma}{\sqrt{(tZ\gamma)^2 +\Delta^2}} \Big)
\;, \label{NUM-F}
\end{eqnarray}
where $E^{\pm}(\gamma) = \pm \sqrt{(tZ\gamma)^2 +\Delta^2} -\mu_{\text{eff}}$ and $\Delta_{\sigma}^2=\Delta^2$.

%%%%%%%%%%%%%%%%%%%%%%%%%%
%%%%%%%%%%%%%%%%%%%%%%%%%%
\section{RESULTS AND DISCUSSIONS}
%%%%%%%%%%%%%%%%%%%%%%%%%%
%%%%%%%%%%%%%%%%%%%%%%%%%%

%%%%%%%%%%%%%%%%%%%%%%%%%%%%%%%%%%%%%%%
\subsection{HALF-FILLED SYSTEM}\label{HALF-FILL}
%%%%%%%%%%%%%%%%%%%%%%%%%%%%%%%%%%%%%%%

In this case, the particle-hole symmetry implies $\mu^{\text{eff}}=0$ and $E^{+}(\gamma) >0$,
and by relation~(\ref{constraint}), Eqs.~\eqref{EPS-F} and \eqref{NUM-F} are simply
\begin{eqnarray}
\epsilon &=& - \frac{t\lambda}{6} I_\epsilon(\lambda) \;, \\
\langle S_{(A/B)\sigma}^z \rangle &=& (-/+) \frac{\,{\rm sgn}({\Delta}_\sigma)\,}{2} I_S(\lambda) \;,
\end{eqnarray}
where $\lambda = tZ/\Delta_{\text{AFM}}$ and
\begin{eqnarray}
I_\epsilon(\lambda) &=& \int_{0}^{3} d\gamma\, D(\gamma) \frac{\gamma^2}{\sqrt{(\gamma \lambda)^2 +1}} \;,\label{Iepsilon}\\
I_S(\lambda) &=& \int_{0}^{3} d\gamma\, D(\gamma) \frac{1}{\sqrt{(\gamma\lambda)^2 +1}} \;. \label{IS}
\end{eqnarray}

\begin{figure}[htb!]
\centering{
\includegraphics[%width=0.8\textwidth,height=0.6\textwidth
scale=0.7]{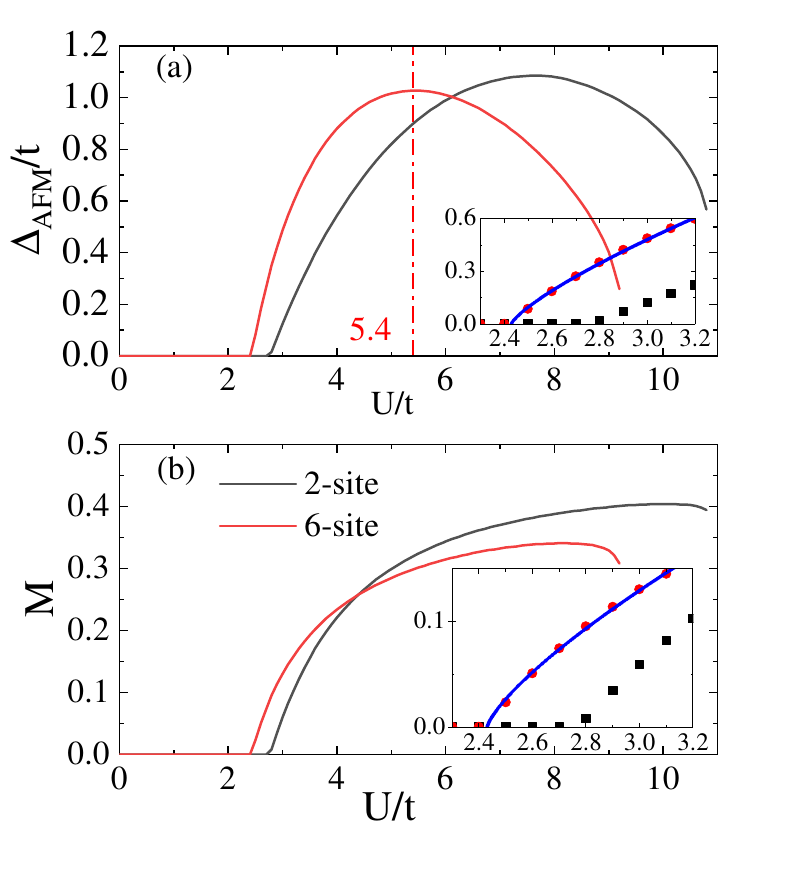}
}
\caption{(a) The AFM energy gap $\Delta_{\text{AFM}}$  and (b) the staggered magnetization $M$  as function of $U$,
where the insets show the same data in the vicinity of the critical coupling
together with the fitting data (blue lines).  The results from two-site cluster approximation are presented as well (black).}
\label{M-DELTA-DOPE00}
\end{figure}
For the half-filled square lattice Hubbard model at $T=0$, the AFM order emerges for any nonzero $U$ because of the
perfect nesting of the free Fermi surface, whereas the honeycomb lattice is known to have a semi-metal phase at
small $U$ due to the low coordination number which allows more fluctuations and an antiferromagnetic phase at large $U$.
It is well established that there is a single direct and continuous phase transition from semi-metal to AFMI
at a finite critical interaction strength $U_{\text{AFM}}$ for the half-filled honeycomb lattice Hubbard
model. This $U_{\text{AFM}}$ from large-scale QMC simulations mainly locates around $U\approx
3.8$~\cite{Sorella_2012,Otsuka_2016,ostmeyer_2021,Raczkowski_2020,Assaad_2013,txma_2018},
whereas the results from various cluster scenarios, such as cluster dynamical impurity approximation, variational
cluster approximation, dynamical cluster approximation and density matrix embedding theory, are strongly cluster
dependent and the $U_{\text{AFM}}$'s are in a wide range of $1.5 \lesssim U \lesssim 4.0$~\cite{seki_2012, Chen_2014,
PhysRevB.87.205127,Yamada_2016,Hassan_2013}. As shown in Fig.~\ref{M-DELTA-DOPE00}, we find that $U_{\text{AFM}}=2.75$
or 2.43 in our two- or six-site cluster approximation, which is larger than that from the Hartree-Fock approximation of
2.235~\cite{Sorella_1992, Raczkowski_2020}, but smaller than those from QMC
simulations~\cite{Otsuka_2016,Assaad_2013,Sorella_2012, ostmeyer_2021,txma_2018}, reflecting the fact that fluctuations
have been incorporated in the six-site cluster, but not enough to give the accurate value. This shortcoming may be
remedied by enlarging the cluster size and strictly dealing with the constraint $S^{z}_{\alpha}=f^{\dagger}_{\alpha}f_{\alpha} -
\tfrac{1}{2}$ locally.  However, the two-site cluster value of $U_{\text{AFM}}$ is larger than that
from the six-site, necessitating more investigations  on the dependence of $U_{\text{AFM}}$ upon the cluster size.
To extract the critical information around $U_{\text{AFM}}$, we fit our self-consistent data from the six-site cluster
using the $U$ dependent form of $M$ and $\Delta_{\text{AFM}}$ that have been verified by QMC
simulations~\cite{Otsuka_2016,Assaad_2013,Sorella_2012, ostmeyer_2021} and density matrix embedding theory~\cite{Chen_2014}
\begin{equation}\label{DELTA-U}
\Delta_{\text{AFM}}/M=\alpha_{1/2}|U-U_{\text{AFM}}|^{\beta_{1/2}}\;,
\end{equation}
and we obtain
\begin{equation}
\Big\{\begin{array}{ll}\alpha_1=0.74021\pm0.00836\;, & \beta_1=0.77086\pm0.01597\;;\\
\alpha_2=0.19739\pm0.000753\;, & \beta_2=0.755\pm0.0049\;.\end{array}
\end{equation}
The critical exponent $\beta_2=0.755\pm0.0049$ for $M$  is very close to $\beta_2=0.75\pm0.06$~\cite{Otsuka_2016}
and 0.79~\cite{Assaad_2013} from the large-scale QMC simulations, and $\beta_2=0.72$ from the
density embedding theory~\cite{Chen_2014}.
{\color{red} It should be mentioned that $\beta_2=0.86546\pm0.01849$
from the two-site approximation is close to that from the six-site one, and both results fall in the ballpark
of the QMC estimates, reflecting the universal aspect of the critical exponent. }
The critical exponent for single particle gap $\Delta_{\text{AFM}}$ is slightly larger than that of $M$~\cite{Assaad_2013,ostmeyer_2021}.

We now expand asymptotically the integrals $I_\epsilon(\lambda)$ and
$ I_S(\lambda) $ defined in Eqs.~\eqref{Iepsilon} and~\eqref{IS} as $\lambda\to\infty$~\cite{NBRAH}:
\begin{equation}
\epsilon = - \frac{t}{6} \Big( 1.574597- 0.448221 \tilde{\Delta}^2 + \frac{4}{3\sqrt{3}\pi} \tilde{\Delta}^3
+ \cdots \Big)\;,
\end{equation}
\begin{eqnarray}
\langle S_{(A/B)\sigma}^z \rangle &=& (-/+) \frac{\,{\rm sgn}({\Delta}_\sigma)\,}{2} \Big(
0.896441 \tilde{\Delta} -\frac{2}{\sqrt{3}\pi} \tilde{\Delta}^2\nonumber\\
&&-0.014 \tilde{\Delta}^3 + \frac{4}{9\sqrt{3}\pi} \tilde{\Delta}^4 + \cdots \Big) \;,
\end{eqnarray}
where $\tilde{\Delta}=\lambda^{-1}=\Delta_{\text{AFM}}/tZ$. The relation between $M$ and $\Delta_{\text{AFM}}$
in the honeycomb lattice around $U= U_{\text{AFM}}$ is established as
\begin{eqnarray}\label{STAG-M-SSPIN}
M &=& \tfrac{1}{2}|\langle S_{A\sigma}^z \rangle-\langle S_{B\sigma}^z \rangle|\nonumber\\
&\approx&\frac{1} {2} \Big(
0.896441 \tilde{\Delta} -\frac{2}{\sqrt{3}\pi} \tilde{\Delta}^2-0.014 \tilde{\Delta}^3
+ \frac{4}{9\sqrt{3}\pi} \tilde{\Delta}^4  \Big) \;.\nonumber\\
\end{eqnarray}
To the leading order, $M \propto  \Delta_{\text{AFM}}$, compared to $M \propto  \Delta_{\text{AFM}}
(\ln{\Delta_{\text{AFM}}})^2$ in the square lattice,
supporting the AFM at small $U$ in the latter case  is driven by the perfect nesting of its free Fermi surface.

\begin{figure*}[htb!]
\centering{
\includegraphics[%width=0.8\textwidth,height=0.6\textwidth
scale=0.62]{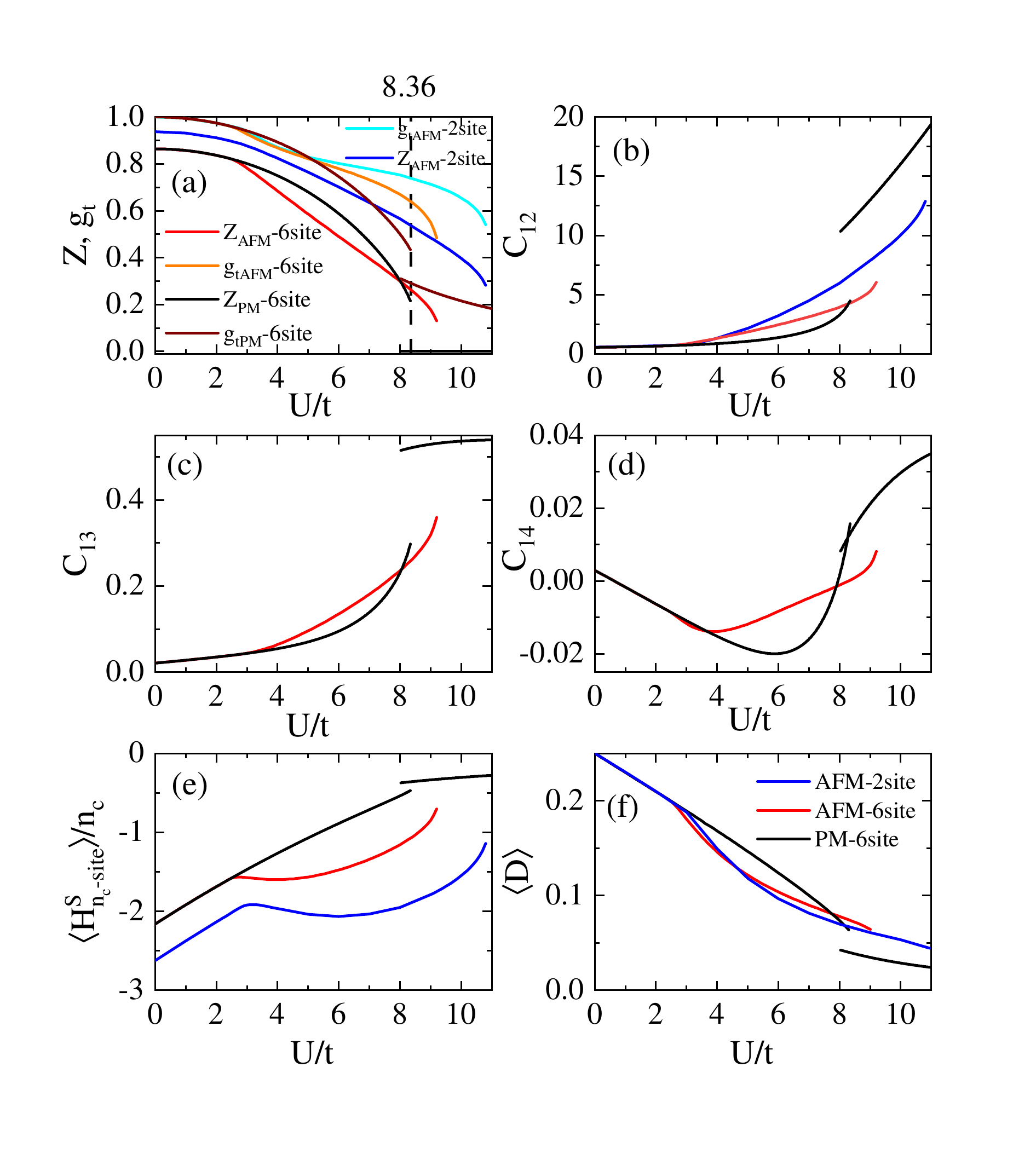}
}
\caption{(a) The quasiparticle weight $Z$ and the generalized Gutzwiller factor $g_t$.  (b)--(d)
The holon-doublon correlators between the nearest neighbors $C_{12}$, the next-nearest neighbors $C_{13}$ and
the  next-next-nearest neighbors $C_{14}$.  (e) The expectation value of the cluster slave-spin Hamiltonian
$\langle H_{n_{c}\text{-site}}^{S}\rangle/n_{c}$.  (f) The double occupancy $\langle D\rangle$ in the AFM state
vs. $U$ obtained by the 2/6-site clusters (blue/red). All quantities in the PM state obtained by 6-site cluster
are black lines.}
\label{dope00-2-6site}
\end{figure*}

On the other hand, $\Delta_{\text{AFM}}$ reaches its maximum around the crossover coupling strength $U_{c}=5.4t$ that
separates the weak- and strong-coupling regimes, which is consistent with the traditional mean-field behavior
$\Delta_{\text{AFM}} \sim U$ at small $U$, and $\Delta_{\text{AFM}} \sim 4 t^{2}/U$ in the large $U$ limit
 supported by the super-exchange mechanism. It ought to be mentioned that $M$ drops abruptly when $U$
is larger than a certain value where the quasi-particle weight happens to drop to zero as shown in
Fig.~\ref{dope00-2-6site}(a), implying that at half-filling, the cluster slave-spin method is incapable
of capturing the crossover between the Hubbard model with finite $U$ and its counterpart in the large $U$
limit---the Heisenberg model, which can be understood from the expression of
$M$ at half-filling
\begin{equation}\label{M-half}
M = \frac{1}{2}|n_{A\sigma} - n_{B\sigma}| = \int_{0}^{3} d\gamma\, D(\gamma)
\frac{|\Delta_\sigma|}{\sqrt{(tZ\gamma)^2 +\Delta^2}}\;,
\end{equation}
where the integration encounters $\frac{0}{0}$ when the AFM energy gap $\Delta_{\text{AFM}}$ and the quasi-particle
residue $Z$ drop to zero simultaneously at large $U$ [See Fig.~\ref{M-DELTA-DOPE00}(a) and \ref{dope00-2-6site}(a)].
However, for a doped system, $Z$ decreases to a constant [Fig.~\ref{SSPINdope002}(a)] to be free from this glitch.

The quasi-particle residue $Z$, the generalized Gutzwiller factor $g_t$~\cite{TohruOgawa, PhysRevB.88.094502,
Lee.Ting-Kuo_2017}, the holon-doublon correlators $C_{ij}$~\cite{Lee.Ting-Kuo_2017} between the nearest neighbors
$C_{12}$,  the next-nearest neighbors $C_{13}$ and the next-next-nearest neighbors $C_{14}$, the ground state energy
of the slave-spin Hamiltonian per site $\langle H_{n_{c}\text{-site}}^{S}\rangle/n_{c}$ with $n_{c}$ being the
cluster size, and the double occupancy $\langle D \rangle$ as function of $U$ at half-filling obtained from two-
and six-site cluster approximations are presented in Fig.~\ref{dope00-2-6site}, where $C_{ij}$ is defined as
\begin{equation}
C_{ij}=\frac{\langle N_{i} D_{j}\rangle-\langle N_{i}\rangle\langle D_{j}\rangle}{\langle
N_{i}\rangle\langle D_{j}\rangle} \;,
\end{equation}
with the holon operator $ N_{i}=(1-n_{i\sigma})(1-n_{i\bar{\sigma}})$ and doublon operator $ D_{j}
=n_{j\sigma}n_{j\bar{\sigma}}$.

The results are as follows:
(i) All quantities in the PM state show  discontinuities and hystereses at the
critical coupling strength $U_{\text{Mott}}=8.36$ for the semi-metal to paramagnetic insulator transition
as the characteristics of the first-order Mott transition in the PM state~\cite{Hassan_2013,Tran_2009}.
(ii) In Fig.~\ref{dope00-2-6site}(a), compared to that in the PM state, $Z$ is largely suppressed as entering
the AFM phase.
(iii) In Fig.~\ref{dope00-2-6site}(b)--(d), $C_{12}$ and $C_{13}$ are positive and increase monotonically with $U$,
signaling that the holon and doublon between the nearest and next-nearest neighbors tend to attract each other,
which is enhanced by the coupling strength. However, $C_{14}$ presents a negative minimum beyond
the AFM transition or as $U$ approaches $U_{\text{Mott}}$ in the PM state,
suggesting that at half-filling, the holon and doublon between the next-next-nearest neighbors attract each other
when $U$ is small or large, while behave repulsively at intermediate $U$.
(iv) In Fig.~\ref{dope00-2-6site}(e), $\langle H_{n_{c}\text{-site}}^{S}\rangle/n_{c}$ in the AFM state is smaller
than that from the PM state, favoring an AFM ground state.
(v) In Fig.~\ref{dope00-2-6site}(f), $\langle D \rangle$ in the PM state decreases linearly with the increasing $U$
when $U \ll U_{\text{Mott}}$~\cite{Vollhardt_1984}, whereas in the AFM state, its slope changes abruptly
as AFM sets in denoting a second-order transition from a semi-metal to an AFMI.

%%%%%%%%%%%%%%%%%%%%%%%%%%%%%%%%%%%%%%%
%%%%%%%%%%%%%%%%%%%%%%%%%%%%%%%%%%%%%%%
\subsection{SYSTEMS WITH FINITE DOPING}\label{FINITEDOPE}
%%%%%%%%%%%%%%%%%%%%%%%%%%%%%%%%%%%%%%%
%%%%%%%%%%%%%%%%%%%%%%%%%%%%%%%%%%%%%%%

\begin{figure*}[htb!]
\centering{
\includegraphics[%width=0.8\textwidth,height=0.8\textwidth
scale=1.21]{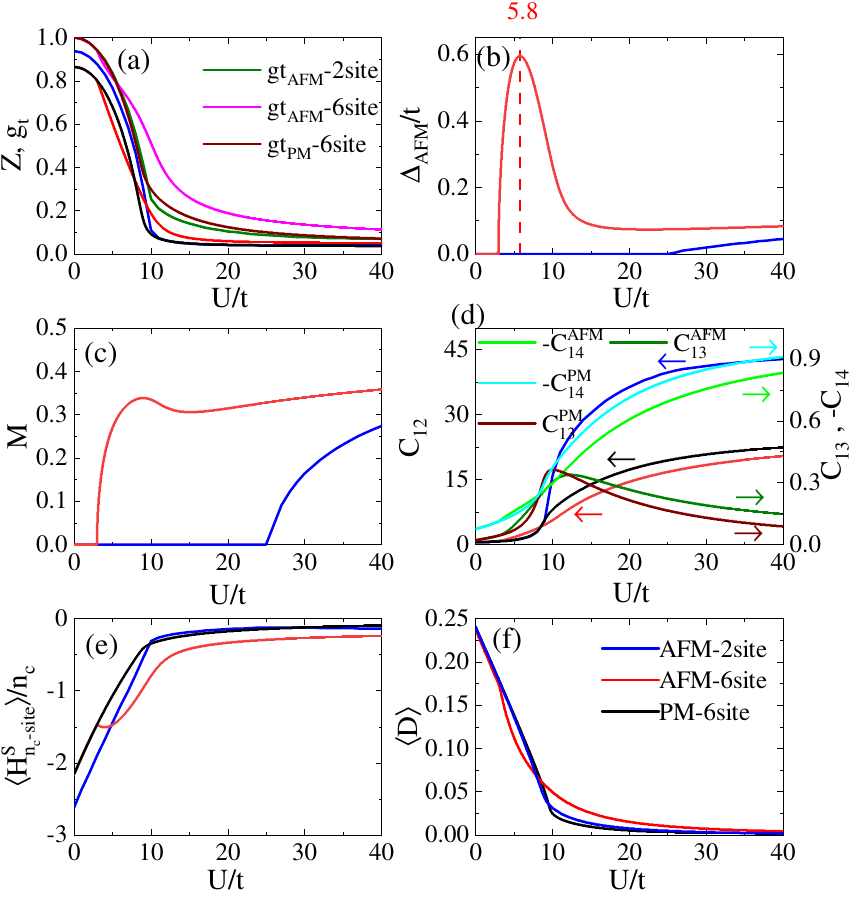}
}
\caption{(a) The quasi-particle weight $Z$ and the generalized Gutzwiller factor $g_t$,
(b) the AFM energy gap $\Delta_{\text{AFM}}/t$,
(c) the staggered magnetization $M$,
(d) the holon-doublon correlators between the nearest neighbors $C_{12}$,  the next-nearest neighbors  $C_{13}$ and
the next-next-nearest neighbors  $C_{14}$,
(e) the expectation value of the cluster slave-spin Hamiltonian $\langle H_{n_{c}\text{-site}}^{S}\rangle/n_{c}$,
and (f) the double occupancy $\langle D\rangle$ as function of $U$ at $\delta=0.02$  in the AFM state obtained by
the two-site (blue) and six-site (red) cluster, as well as the PM state (black) by six-site cluster.}
\label{SSPINdope002}
\end{figure*}
\begin{figure}[htb!]
\centering{
\includegraphics[width=0.47\textwidth,scale=0.7]{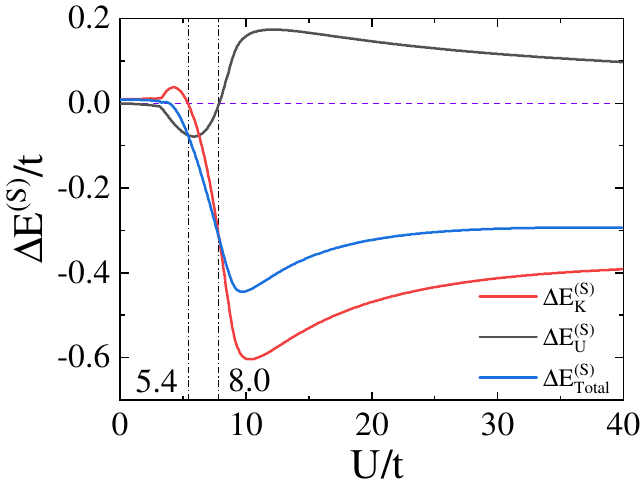}
}
\caption{The difference of the kinetic energy $\Delta E_{K}^{(S)}$ (red), interaction potential $\Delta
E_{U}^{(S)}$ (black), and their summation $\Delta E_{\text{Total}}^{(S)}$ (blue) of the six-site cluster
slave-spin Hamiltonian Eq.~\eqref{H_K} between the AFM and PM states as function of $U$ at $\delta=0.02$.}
\label{D-EK-DOPE002}
\end{figure}

In Fig.~\ref{SSPINdope002}, we plot $Z$,  $g_t$, $\Delta_{\text{AFM}}$, $M$, $C_{12/3/4}$, $\langle
H_{n_{c}\text{-site}}^{S}\rangle/n_{c}$, $\langle D \rangle$
as function of $U$ at $\delta=0.02$ obtained from two- and six-site cluster approximations to further compare the results
from these two slave-spin clusters.
In Fig.~\ref{SSPINdope002}(a), the quasi-particle residue from six-site cluster is much smaller than that from two-site
because of more quantum fluctuations, and becomes flattened when $U>U_{\text{Mott}}$.
In Fig.~\ref{SSPINdope002}(b), the critical coupling strength for AFM transition from six-site cluster is
$U_{\text{AFM}}=3.0$, while that from two-site locates at $U_{\text{AFM}}=25.0$, which indicates that two-site cluster
is inadequate to describe the AFM transition in the honeycomb lattice because it keeps no track of the lattice symmetry.
In Fig.~\ref{SSPINdope002}(d), $C_{12}$ increases slowly when
$U<U_{\text{Mott}}$, then rises dramatically as $U$ approaches $U_{\text{Mott}}$, and finally grows progressively as
$U$ goes to infinity; while $C_{13}$ shows a maximum near $U_{\text{Mott}}$, the reason for which is that the hopping
probability between the next-nearest neighbors falls faster than the one between the nearest neighbors when $U$ is
increased as demonstrated in our previous work on a square lattice~\cite{zeng2021}. Unlike  $C_{12}$ and $C_{13}$, both
positive for all $U$'s, $C_{14}$ is negative at $\delta=0.02$ and its magnitude grows monotonically with the coupling strength,
indicating that the holon and doublon between the next-next-nearest neighbors repulse each other,
whose tendency is strengthened as $U$ increases. In Fig.~\ref{SSPINdope002}(e), as shown by the blue line with $0<U<25$
(where the system within two-site approximation is in the PM state.) and the black line, the difference of
$\langle H_{n_{c}\text{-site}}^{S}\rangle/n_{c}$ in PM state between two- and six-site cluster approximations
is much smaller when $U>U_{\text{Mott}}$, denoting that the cluster size's effect on the properties of the
system in the PM state is less important at large $U$ as the system becomes more localized, where the
inter-site fluctuations are much weaker in contrast to the weak-coupling limit.  In Fig.~\ref{SSPINdope002}(f),
there exists an inflection in $\langle D \rangle$ in the PM state around $U\approx10t$, meaning that the first-order
Mott transition at half-filling turns into a continuous crossover at finite dopings.
For $U<8.0$, the double occupancy in the AFM state is smaller than that in the PM state while the opposite is true
for $U>8.0$, bespeaking that the AFM at small $U$ is triggered by the interaction potential gain while that in the
large $U$ limit is not driven by this mechanism. This picture can also be seen in Fig.~\ref{D-EK-DOPE002},
where exists a region ($5.4<U<8.0$) with $\Delta E_{U}^{(S)}<0$ and $\Delta E_{K}^{(S)}<0$, signaling that the AFM
in this region is supported by both the kinetic energy and interaction potential gain.

\begin{figure*}%[htb!]
\centering{
\includegraphics[%width=0.8\textwidth,height=0.8\textwidth
scale=0.54]{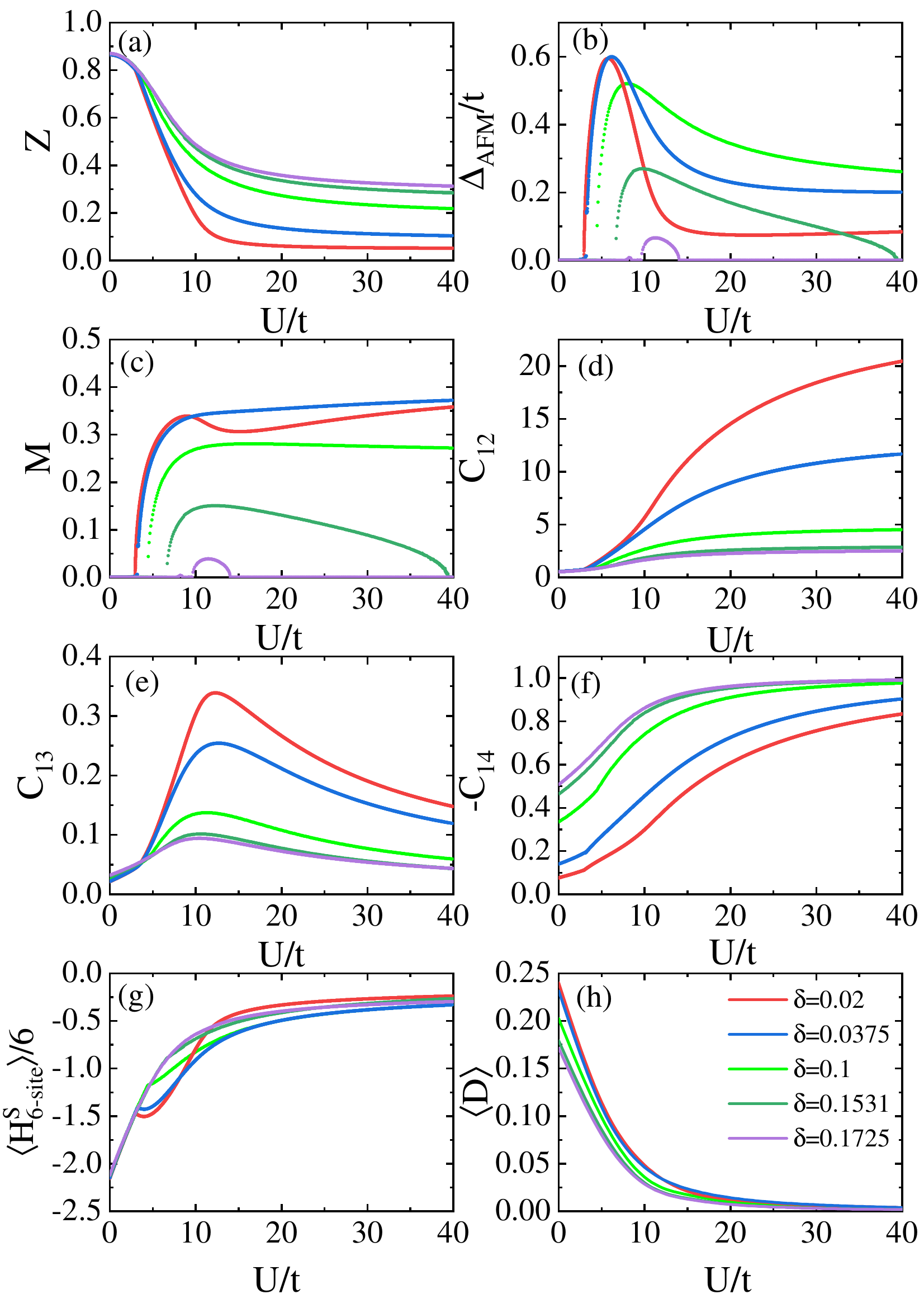}
}
\caption{(a) The quasiparticle weight $Z$, (b) the AFM energy gap $\Delta_{\text{AFM}}/t$, (c) the staggered
magnetization $M$, (d)--(f) the holon-doublon correlator between the nearest neighbors $C_{12}$,  the next-nearest
neighbors  $C_{13}$  and  the next-next-nearest neighbors $C_{14}$,  (g) the expectation value of the cluster
slave-spin Hamiltonian,  and (h) the double occupancy $\langle D\rangle$ as function of $U$  at a series of
doping concentrations $\delta=$ 0.02 (red), 0.0375 (blue), 0.1 (green), 0.1531 (dark green), 0.1725 (violet)
in the AFM state obtained by the six-site cluster.}
\label{dope-vari-6site}
\end{figure*}
The dependence of the quantities discussed above upon the interaction strength at various dopings of
$\delta=0.02,\,0.0375,\,0.1,\,0.1531,\,0.1725$ obtained through six-site cluster approximation are presented in
Fig.~\ref{dope-vari-6site}, where $\delta=0.1531$ is the critical doping for the AFM to PM phase transition
at $U=40$, and $\delta=0.1725$ is the maximum of $\delta_{M}$, i.e., the boundary between the AFM and PM state
[see Figs.~\ref{M-DOPE-U} and \ref{DOPE-U-PHASES}].
The following results are concluded:
(i) In Fig.~\ref{dope-vari-6site}(a), the increasing $Z$ with $\delta$ for all coupling strengths suggests that
the system with the increasing doping tends to be metallic. When $U>U_{\text{Mott}}$, the quasi-particle residues
decrease progressively to constants,  manifesting that the properties in the AFM state are controlled by the
underlying Mott transition.
(ii) In Fig.~\ref{dope-vari-6site}(b), $\Delta_{\text{AFM}}$'s at all dopings exhibit a maximum around the
crossover coupling strength $U_{c}(\delta)$ that grows with $\delta$.
(iii)  In Fig.~\ref{dope-vari-6site}(d),  $C_{12}$ increases monotonically with $U$ at all dopings and diminishes
as $\delta$ goes up, which makes it eligible to be an indicator of the magnitude of correlations.
(iv) In Fig.~\ref{dope-vari-6site}(g),  compared to the PM state, $\langle H_{\text{6-site}}^{S}\rangle/6$ is
suppressed dramatically as soon as AFM sets in and this effect is weakened by the increasing $\delta$.

\begin{figure}[htb!]
\centering{
\includegraphics[width=0.47\textwidth,scale=0.7]{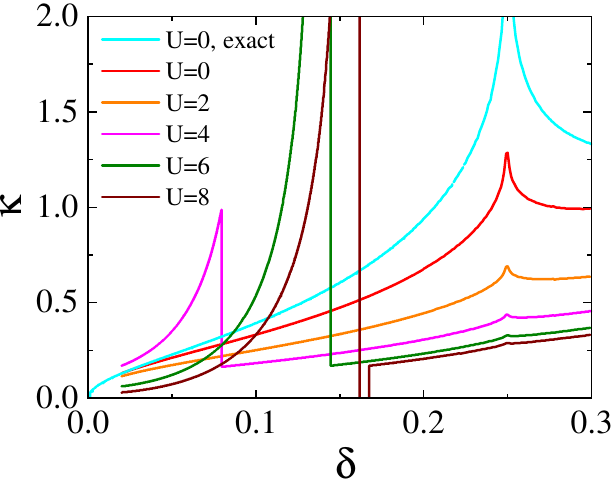}
}
\caption{The compressibility $\kappa$ obtained from the six-site cluster approximation as function of $\delta$
at $U=0,\,2,\,4,\,6,\,8$ (red, orange, magenta, olive, brown, respectively), and the exact one at $U=0$
calculated through Eq.~\eqref{kapa_U0} (cyan).}
\label{COMPRESSIBILITY}
\end{figure}

The compressibility of the system is defined as $\kappa = n^{-2} \partial n/\partial \mu$. At $U=0$, it is calculated
by using the non-interacting DOS $D(\gamma)$, Eq.~\eqref{density}, via
\begin{eqnarray}\label{kapa_U0}
\kappa(\mu) = \Bigg\{ \begin{array}{ll}
\frac{N(|\mu|)}{[\frac{3}{4} +\int_{-1}^{\mu} N(|\gamma|) d \gamma]^2} \;, & \;\;  (0<\delta<\frac{1}{4}) \\
\frac{\tilde{N}(|\mu|)}{[\int_{-3}^{\mu} \tilde{N}(|\gamma|) d \gamma]^2} \;, & \;\; (\frac{1}{4}<\delta<1)
\end{array}
\end{eqnarray}
which is proportional to the free DOS.
For $U>0$, $\kappa$ should evolve simultaneously with the quasi-particle DOS which makes it adequate to indicate
the dependence of this quantity upon interaction. The $\kappa$'s as function of $\delta$ at $U=0,\,2,\,4,\,6$,
and 8 are plotted in Fig.~\ref{COMPRESSIBILITY}. For $U<U_{\text{AFM}}$, the compressibility near the
van Hove singularity is suppressed most drastically by interaction,
while that at low energy (near the Dirac points) remains very close to the non-interacting one, reflecting
that the Dirac cone structure is very robust,
and the DOS of the quasi-particles is transferred away from the van Hove singularity as $U$ increases.
Furthermore, at $U=4,\,6$, there exhibit a discontinuity at $\delta=\delta_{M}$  where the AFM-to-PM phase transition
occurs, and the one-sided peak of $\kappa$ as $\delta$ approaches $\delta_{M}$ manifests that the system now is
an itinerant AFM metal. However, at $U=8$, there exist two consecutive discontinuities: (i) between positively
and negatively divergent $\kappa$; (ii) between negatively divergent and positive small $\kappa$.

%%%%%%%%%%%%%%%%%%%%%%
%%%%%%%%%%%%%%%%%%%%%%
\section{PHASE DIAGRAM}\label{PHASES}
%%%%%%%%%%%%%%%%%%%%%%
%%%%%%%%%%%%%%%%%%%%%%

\begin{figure}[htb!]
\centering{
\includegraphics[width=0.47\textwidth,scale=0.7]{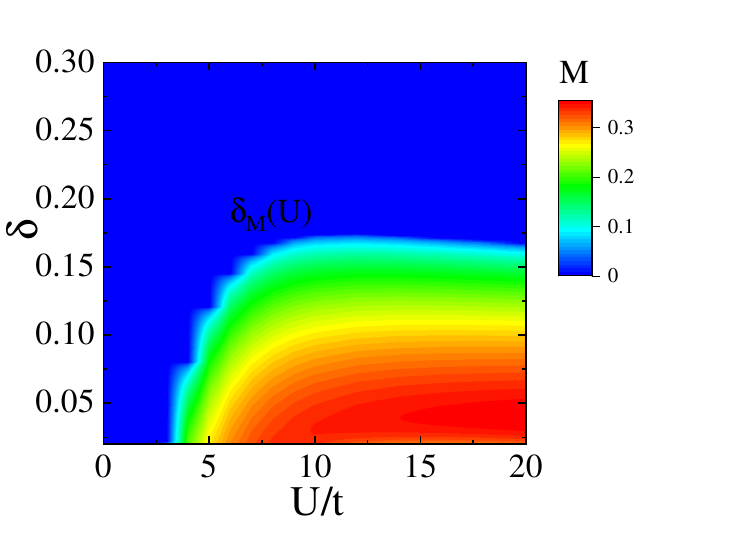}
}
\caption{The staggered magnetization $M$  as function of $U$ and $\delta$ obtained by six-site cluster,
where $\delta_{M} (U)$ is the phase boundary between the AFM and PM states.}
\label{M-DOPE-U}
\end{figure}

The staggered magnetization $M$ with $U$ and $\delta$ being its parameters is plotted in Fig.~\ref{M-DOPE-U},
where the phase boundary  between the AFM and PM states is delineated by $\delta_{M}(U)$. Obviously, $M$ maximizes
at small dopings and large couplings.  The phase boundary $\delta_{M}(U)$ shows a nonmonotonic behavior upon $U$
that may be connected to the crossover of $\Delta_{\text{AFM}}$ as $U$ increases. We also notice that $M$ saturates
when $U>U_{\text{Mott}}$ at small dopings, reflecting that the physical properties in the AFM state are dominated
by the underlying Mott transition in the half-filled PM state.

\begin{figure}[htb!]
\centering{
\includegraphics[width=0.47\textwidth,scale=0.7]{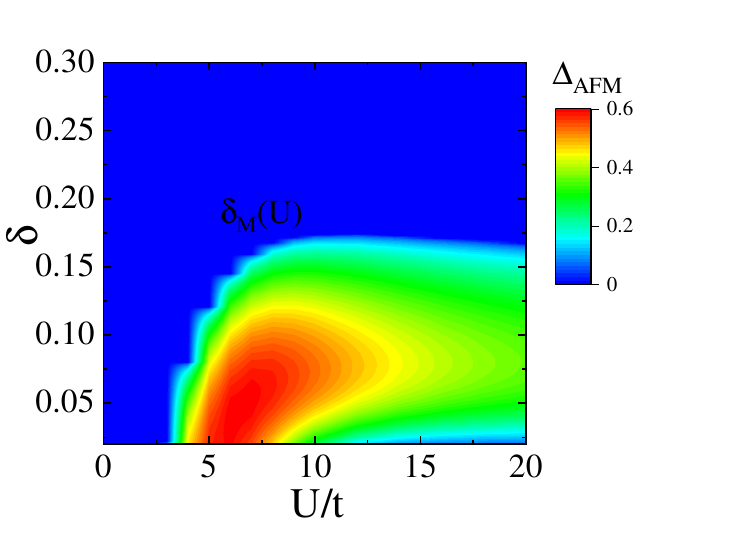}
}
\caption{The AFM gap $\Delta_{\text{AFM}}$  as function of $U$ and $\delta$ obtained by six-site cluster
with $\delta_{M} (U)$ separating the AFM and PM states.}
\label{DELTA-DOPE-U}
\end{figure}

\begin{figure}[htb!]
\centering{
\includegraphics[width=0.47\textwidth,scale=0.7]{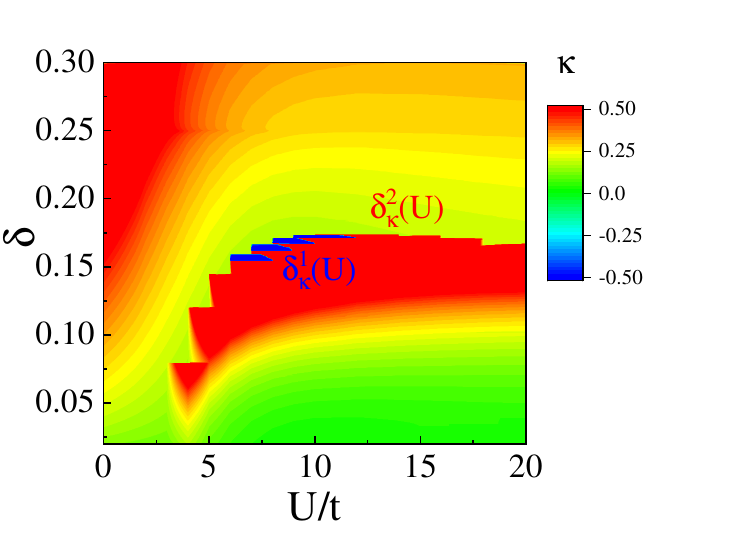}
}
\caption{The compressibility $\kappa$  as function of $U$ and $\delta$ obtained by six-site cluster.
The staircase is an artifact because of discrete $U$'s adopted to calculate $\kappa$,
i.e., $\Delta U=t$
when $U\le 10t$ and $\Delta U=2t$ when $10t<U\le20t$, which can only be eliminated in the $\Delta U\to 0$ limit.
The blue region between $\delta_{\kappa}^{1} (U)$ and $\delta_{\kappa}^{2} (U)$ is characterized by $\kappa<0$,
where $\delta_{\kappa}^{1} (U)$ and $\delta_{\kappa}^{2} (U)$ are delineated by the midpoints of the blue and red
steps, respectively. The discontinuities in $\kappa$ at these two phase boundaries are reflected
in the color jumpings. }
\label{KAPPA-DOPE-U}
\end{figure}

\begin{figure}[htb!]
\centering{
\includegraphics[width=0.47\textwidth,scale=0.7]{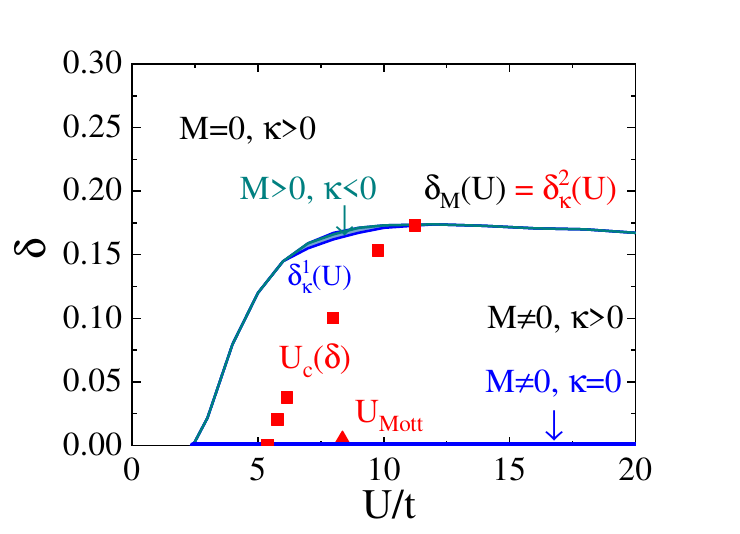}
}
\caption{The $U$-$\delta$ phase diagram of the honeycomb lattice Hubbard model within the six-site
cluster scheme. The critical coupling $U_{\text{Mott}} = 8.36t$ for the Mott transition in the half-filled PM state
is marked by the red triangle. The crossover coupling $U_{c}(\delta)$ in the AFM state
at $\delta =$ 0, 0.02, 0.0375, 0.1, 0.1531 and 0.1725 are symbolled by red squares, at which  $\Delta_{\text{AFM}}$
reaches its maximum. The half-filling case with $U>2.43t$ is highlighted by the heavy blue line,
in which the system is an AFM insulator with $\kappa=0$.
}
\label{DOPE-U-PHASES}
\end{figure}

The AFM energy gap $\Delta_{\text{AFM}}$ in the same parameter space is plotted in Fig.~\ref{DELTA-DOPE-U},
where the phase boundary $\delta_{M}(U)$ still holds, denoting that there is no intermediate states before
the semi-metal-to-AFMI transition occurs. An overall crossover between the weak- and strong-coupling regimes
can be observed in $\Delta_{\text{AFM}}$ when $U$ grows, at which $\Delta_{\text{AFM}}$ reaches its maximum, and the
coupling $U_{c}$ for this crossover is highly $\delta$-dependent, in contrast to that in the square lattice case~\cite{zeng2021}.
For $U>U_{c}$, the maximum of $\Delta_{\text{AFM}}$ occurs at $\delta \sim 0.075$, leading to
an interesting vertical re-entrance behavior as $\delta$ increases, same as the square lattice case~\cite{zeng2021}.

Combining Figs.~\ref{M-DOPE-U}, \ref{DELTA-DOPE-U}, and \ref{KAPPA-DOPE-U}, an overall phase diagram in the
$U$-$\delta$ plane emerges in Fig.~\ref{DOPE-U-PHASES}. In contrast to the square lattice case~\cite{zeng2021},
the crossover $U_{c}$ in the AFM state at which $\Delta_{\text{AFM}}$ is maximized, symbolled by the red squares,
is shown to be highly $\delta$-dependent. On the other hand, $U_{c}$ at $\delta=0$ is smaller than $U_{\text{Mott}}$
(red triangle), implying that at half-filling, the coupling strength separating the weak- and
strong-coupling regimes is suppressed by long-range AFM correlations~\cite{zeng2021}. The blue region in this figure
enclosed by $\delta^1_{\kappa}(U)$ and $\delta^2_{\kappa}(U)$ is characterized by $M\neq 0$ and $\kappa<0$ with
$\delta^1_{\kappa}(U)$ being the phase boundary between $M\neq 0$, $\kappa>0$ and $M\neq 0$, $\kappa<0$,
and $\delta^2_{\kappa}(U)$ between $M\neq 0$, $\kappa<0$ and $M= 0$, $\kappa>0$. The region with $\kappa<0$
is extremely small compared to the square lattice Hubbard model~\cite{zeng2021}, and exists only in
the vicinity of the phase boundary between the AFM and PM state and at intermediate $U$.
It should be noted that the phase diagram has been greatly improved from the two-site to six-site schemes,
since in the former case $U_{\text{AFM}}$ jumps from $2.75t$ at $\delta = 0$ to $25.5t$ at $\delta = 0.02$, while in the latter
it almost continuously from $2.43t$ to $2.97t$.

%%%%%%%%%%%%%%%%%%%%%
%%%%%%%%%%%%%%%%%%%%%
\section{CONCLUSION}\label{conclude}
%%%%%%%%%%%%%%%%%%%%%
%%%%%%%%%%%%%%%%%%%%%

We have exploited the cluster slave-spin method to explore extensively the ground state properties of the one-band
honeycomb lattice Hubbard model with $U$ and $\delta$ as its parameters. At half-filling, the first-order semi-metal
to insulator Mott transition in the PM state is revealed, characterized by discontinuities and hystereses in all
quantities at $U_{\text{Mott}} =8.36t$~\cite{Hassan_2013, Chen_2014,Tran_2009}. In the AFM state, a single
direct and continuous phase transition between PM semi-metal and AFMI at $U_{\text{AFM}}=2.43t$ is substantiated,
which belongs to the Gross-Neveu-Yukawa universality class~\cite{Otsuka_2016, Raczkowski_2020, Assaad_2013, Paiva_2005,
Yamada_2016, Sorella_1992, Sorella_2012, Ostmeyer_2020,Hassan_2013}, precluding the existence of intermediate phases
such as a spin liquid state. At finite dopings, an extended crossover is discovered between the weak- and strong-coupling
regimes in the AFM state at which the AFM energy gap $\Delta_{\text{AFM}}$ reaches its maximum,
and the AFM within this crossover is driven by both the kinetic energy and interaction potential gain.
The interaction $U_{c}$ for this crossover is shown to be highly $\delta$-dependent,
in contrast to the square lattice system where $U_{c}$ remains almost unchanged
with large dopings~\cite{zeng2021}.

Moreover, for the half-filled system, by analytically calculating the relation between $M$ and $\Delta_{\text{AFM}}$
in the vicinity of PM semi-metal to AFM insulator transition, Eq.~\eqref{STAG-M-SSPIN}, we found that to the
leading order, $M$ is linearly dependent on $\Delta_{\text{AFM}}$, compared to the square lattice result that is
proportional to $\Delta_{\text{AFM}}(\ln{\Delta_{\text{AFM}}})^2$ (Ref. \onlinecite{zeng2021}). This difference is
consistent with the vanishing non-interacting DOS at Dirac points in the honeycomb lattice, in contrast to the van Hove
singularity of the free electron DOS at the Fermi surface in a half-filled square lattice.

Finally, an overall phase diagram in the $U$-$\delta$  plane is presented in Fig.~\ref{DOPE-U-PHASES},
the phase boundary $\delta_{M}(U)$ separating the AFM and PM phases shows a nonmonotonic behavior
with the increasing $U$, which is consistent with the crossover behavior of $\Delta_{\text{AFM}}$.
The phase boundary between the AFM metal with $\kappa>0$ and the AFM insulator with $\kappa = 0 $
locates exactly at $\delta = 0$. The region with $\kappa<0$ only exists in the vicinity of the phase boundary
between the AFM and PM state and at intermediate coupling strengths, whose area is extremely small compared
to the counterpart in the square lattice Hubbard model~\cite{zeng2021}.

It is worth mentioning that though lacking available data from the previous studies to verify
our results at finite dopings, we corroborate that there is no intermediate states such as a spin liquid between
the PM semi-metal and AFMI phases at half-filling~\cite{Otsuka_2016, Raczkowski_2020, Assaad_2013, Paiva_2005,
Yamada_2016, Sorella_1992, Sorella_2012, Ostmeyer_2020,Hassan_2013}, and the critical transition exponent of staggered
magnetization between these two states is quite close to those from large-scale QMC simulations~\cite{Sorella_2012,Assaad_2013,
Otsuka_2016,ostmeyer_2021} and DMET calculations~\cite{Chen_2014}, which could well justify our calculations.
 We would like to mention that the interesting physics in the Hubbard model on a honeycomb lattice could be connected with
 the properties of graphen-based material~\cite{PhysRevLett.97.146401,RevModPhys.81.109,txma_2018,
 doi:10.1063/1.3485059,*PhysRevB.84.121410,*PhysRevB.90.245114}, and also the optical lattice systems for ultracold
 atoms~\cite{PhysRevLett.115.115303}. The ionic Hubbard model with ultracold fermions based on the honeycomb lattice
 has been realized where the transition from metal to charge density wave has been observed, and our theoretical
 prediction is consistent with the experimental results at the limit of stagger potential equal to zero.
We hope our full phase diagram in the parameter space of on-site interaction and doping may simulate further
experimental detection on graphene-based material or optical lattice systems for ultracold atoms.

%%%%%%%%%%%%%%%%%%%%%
%%%%%%%%%%%%%%%%%%%%%
\begin{acknowledgments}
%%%%%%%%%%%%%%%%%%%%%
%%%%%%%%%%%%%%%%%%%%%

We thank Shiping Feng, Xiong Fan and Yu Ni for many helpful discussions.
One of authors (MHZ) would like to acknowledge the beneficial
communications with Rong Yu.  This work was supported by NSFC
(Nos.~11974049 and 11774033), Beijing Natural Science
Foundation (No.~1192011), and the HSCC program of
Beijing Normal University.

\end{acknowledgments}

\bibliography{HEXA-LITERATURE-LIB}

\end{document}